\documentclass[checkin,pre]{revtex4}

\usepackage[dvips]{graphicx}
\usepackage{amssymb}

\begin{document}


\title{Microscopic and Macroscopic Stress \\
with Gravitational and  Rotational Forces.}

\author{Wm. G. Hoover and Carol G. Hoover \\
Ruby Valley Research Institute \\ Highway Contract 60,
Box 598, Ruby Valley 89833, NV USA \\
     }

\author{James F. Lutsko \\
Physics Department CP 231 \\
Universit\'e Libre de Bruxelles \\
Blvd. du Triomphe, 1050 Brussels, Belgium \\
     }




\vskip 0.5cm

\begin{abstract}

Many recent papers have questioned Irving and Kirkwood's atomistic expression
for stress.  In Irving and Kirkwood's approach both interatomic forces and
atomic velocities contribute to stress.  It is the velocity-dependent
part that has been disputed. To help clarify this situation we investigate
(i) a fluid in a gravitational field and (ii) a steadily rotating solid.  For
both problems we choose conditions where the two stress contributions,
potential and kinetic, are significant.  The analytic force-balance solutions
of both these problems agree very well with a smooth-particle interpretation
of the atomistic Irving-Kirkwood stress tensor.

\end{abstract}

\maketitle

\section{Introduction}

In 2003 Zhou published his lengthy and detailed ``New Look at the Atomic Level
Virial Stress'' in the Proceedings of the Royal Society of London\cite{b1}.
He criticized the usual Irving-Kirkwood virial expression\cite{b2} for the
pressure tensor $P$ as a sum of potential and kinetic terms.  The pressure
tensor is the same thing as the comoving corotating momentum flux, and is
also minus the stress tensor, $\sigma \equiv -P$. The detailed microscopic
Irving-Kirkwood approach has been used
for more than 50 years in the interpretation of atomistic molecular dynamics
simulations.\cite{b3,b4,b5,b6}  Averaged over a homogeneous periodic volume
$V$
the Irving-Kirkwood expression for the pressure tensor gives:
$$
-\sigma V \equiv PV = P^\Phi V + P^K V =
\sum_{i<j} (Fr)_{ij} + \sum_i (pp/m)_i \ .
$$
Here $F_{ij}$ is the force (for simplicity we assume a pairwise-additive
potential) exerted on Particle $i$ by Particle $j$, where the
vector from $j$ to $i$ is $r_{ij}$.  Particle $i$, at location $r_i$ with
mass $m_i$ and momentum $p_i$, obeys Newton's equation of motion,
$$
m_i\ddot r_i \equiv F^{\rm ext}_i + \sum_{j\neq i}F_{ij} \ ; \
F_{ij} = -\nabla _i \phi(|r_{ij}|) \ ; \ \Phi \equiv \sum _{i<j}\phi_{ij} \ .
$$
Zhou stated that only the tensor force sum,
$\sum (Fr)_{ij} \equiv \sum F_{ij}r_{ij}$, contributes to the stress,
while the tensor momentum sum,
$\sum (pp/m)_i \equiv \sum (p_ip_i/m_i)$, does not.

This idea -- including the forces but not the momenta -- is not quite so
outlandish as it seems.  In solids, where the longtime average of the
particle location is a sensible quantity, the virial theorem {\em can}
be written in a similar tensor form omitting the momenta:
$$
\langle PV \rangle = \sum_{i<j} \langle (FR)_{ij}\rangle  \ ;
\ R_i \equiv \langle r_i \rangle \ .
$$
This form is derived in Section II.C of Reference 4.  We use angular brackets
here to indicate longtime averages.  In situations including external
forces the tensor force sum must also include either $(F^{\rm ext}r)_i$ or
$(F^{\rm ext}R)_i$.

Subramaniyan and Sun\cite{b7} tested Zhou's ideas with molecular dynamics,
heating a model atomistic solid subject to a variety of external boundary
conditions on the particle coordinates.  Their simulations showed that
only the full Irving-Kirkwood pressure tensor, potential plus kinetic, was
consistent with macroscopic thermodynamics.  Liu and Qiu\cite{b8} recently
provided a
useful list of references supporting both sides of the question.  In addition
they suggest that the Zhou prescription is correct provided that external
fields and rotation are not involved.  Here we explore those latter two
conditions separately and explicitly, showing that both [1] an external field
(gravity) and [2] condensed-phase rotation can be analyzed properly with the
Irving-Kirkwood pressure tensor, in a way compatible with macroscopic
continuum mechanics.  This suggests that the original Irving-Kirkwood
approach is more generally useful than is Zhou's suggested modification of
it.

In order to compute continuous differentiable field variables (density,
velocity, energy, stress, heat flux, ... ) from atomistic molecular
dynamics simulations, for comparison to corresponding fields generated
by continuum mechanics solutions, we recommend the use of ``smooth-particle''
averages.
These correspond to smearing individual particle properties over a spatial
region of size $h$, the range of the smooth-particle weighting function,
as is described in a recent text\cite{b9}, summarized in Section II, and
applied in Section III.

Because the derivation of the pressure tensor is familiar, and applies both
at and away from equilibrium\cite{b4,b5} we do not repeat that here.
Instead, in Sections III and IV, we describe and study two specially
instructive problems involving gravitational and rotational forces.  We
reserve our conclusions and closing remarks for Section V.

\section{Smooth-Particle Averages of Atomistic Properties}

Irving and Kirkwood chose to localize {\em particle} properties {\em at} the
particle locations using delta functions.  Though this is convenient for
formal analyses, and even natural for mass and momentum, a smoothed or
smeared-out particle contribution to potential energy and to fluxes often
simplifies comparisons with continuum mechanics.  The smeared approach can
provide field variables with two continuous spatial derivatives, as we show
below.

Because ``action at a distance'' makes
the exact location of momentum and energy fluxes ambiguous we choose to smear
out particle contributions within a spatial region somewhat larger in extent
than the spacing between particles.  We use a local weight function with a
range $h$, $w(r,h)$ to convert particle properties to continuum field
properties.  Consider, for example, the density $\rho $ and the velocity $v$
in a fluid or solid composed of particles with individual masses and
velocities $\{ m_i,v_i\} $.  In the smooth-particle approach\cite{b9,b10}
field variables, such as the density and velocity at the point $r$, are
{\em defined} as $h$-dependent (range-dependent) sums of nearby particle
contributions:
$$
\rho(r) \equiv \sum _jm_jw(|r-r_j|) \ .
$$
$$
\rho(r)v(r) \equiv \sum _jm_jv_jw(|r-r_j|) \ .
$$
The sums include all particles within a distance $h$ of point $r$.  A good
feature of this approach is that these definitions of density and velocity
satisfy the continuity equation, $\dot \rho/\rho \equiv - \nabla \cdot v$,
exactly.  Here, as is usual, the dot indicates a comoving time derivative
following the motion.

Lucy was one of the inventors of the smooth-particle approach\cite{b10}.  For
convenience we use his form for the weighting function in all of our
smooth-particle sums,
$$
w_{\rm Lucy}^D(|r|<h) = C_D(1 - 6x^2 + 8x^3 - 3x^4) \ ; \ x = |r|/h.
$$
This form has two continuous derivatives everywhere.
The normalizing prefactor $C_D$ depends on the dimensionality $D$,
$$
C_1 = (5/4h) \ ; \ C_2 = (5/\pi h^2) \ ; \ C_3 = (105/16\pi h^3) \ .
$$
$C$ is chosen so that the spatial integral of the weight function is unity:
$$
\int_0^h w^1(r)2dr = \int_0^h w^2(r)2\pi rdr  =
\int_0^h w^3(r)4\pi r^2dr \equiv 1 \ .
$$

Lucy's polynomial form is the simplest normalized weight function with
a maximum value at the origin and two continuous derivatives everywhere.
In the following section, where we consider the
mechanical equilibrium of a two-dimensional fluid in a one-dimensional
gravitational field, we compute average values of the pressure tensor
using the one-dimensional form of Lucy's weight function.

\section{Gravitational Equilibration}

Gravitational equilibration is a problem in which both the potential
and kinetic contributions to stress can play a r\^ole.  Where a constant
gravitational acceleration acts downward in $y$, the simple
force balance equation for mechanical equilibrium is,
$$
dP/dy = (dP/d\rho )(d\rho /dy) = - \rho g \ .
$$
The stationary density profile, $\rho (y)$, can be found provided
that the dependence of pressure $P$ on the density $\rho$ is known.
As a simple example problem, chosen to highlight the kinetic and potential
contributions to the virial, we choose to study the molecular dynamics
of an atomistic system which closely approximates the isothermal fluid
equation of state
$$
P(\rho ,T) = (\rho^2/2) + \rho T \ ; \ T \equiv 1 \ .
$$
This equation of state closely corresponds to the virial equation of
state for two-dimensional particles of unit mass at unit temperature
interacting with a ``Cusp'' potential chosen to have a spatial integral
of unity:
$$
\phi_{\rm Cusp}(r<h) = (10/\pi h^2)(1 - x)^3 \ ; \ x = |r|/h \ ; \
$$

$$
\longrightarrow \int_0^h2\pi r\phi _{\rm Cusp}(r)dr \equiv 1 \ .
$$

$$
\langle p_x^2/m \rangle = \langle p_y^2/m \rangle = kT \equiv 1 \ .
$$

We use this cusp interaction for the interparticle forces because the
model closely corresponds to the simple and useful thermodynamic equation
of state given above.  We choose the range of the Cusp pair potential $h=3$, so
that the deviation of the potential part of the pressure tensor from
that macroscopic equation of state is of order one percent.

For periodic two-dimensional systems the virial-theorem expression for
the potential part of the pressure tensor can be expressed in terms of
sums over all $N(N-1)/2$ pairs of interacting particles\cite{b4,b5}.
For a hydrostatic fluid,
where $P^\Phi_{xx}$ and $P^\Phi_{yy}$ are each equal to the potential
part $P^\Phi $ of the hydrostatic pressure $P$, we have:
$$
P^\Phi_{xx}V = \sum (xF_x)_{i<j} = P^\Phi_{yy}V = \sum (yF_y)_{i<j} =
(1/2)\sum (F \cdot r)_{i<j} = P^\Phi V \ .
$$
For a completely random distribution of particles in the volume $V$ the
potential part of the pressure is then given by a force integral.
The integral can be related to the integral of the pair potential using
integration by parts.  With our particular choice of pair potential $\phi $,
with an integral of unity, and particle mass, unity, the resulting
hydrostatic pressure is simply half the square of the density:
$$
P^\Phi V = (1/2)\sum (F \cdot r)_{i<j} \simeq
$$
$$
-[N(N-1)/(4V)]\int_0^h2\pi r^2\phi^\prime dr \equiv
$$
$$
+[N(N-1)/(2V)]\int_0^h2\pi r\phi dr \equiv N(N-1)/(2V)
$$
$$
\simeq N\rho /2
\longrightarrow P^\Phi \simeq (1/2)\rho ^2 \ .
$$

A snapshot from an isokinetic (constant kinetic temperature) simulation
appears in {\bf Figure 1}.

\begin{figure}
\includegraphics[height=12cm,width=10cm,angle=-0]{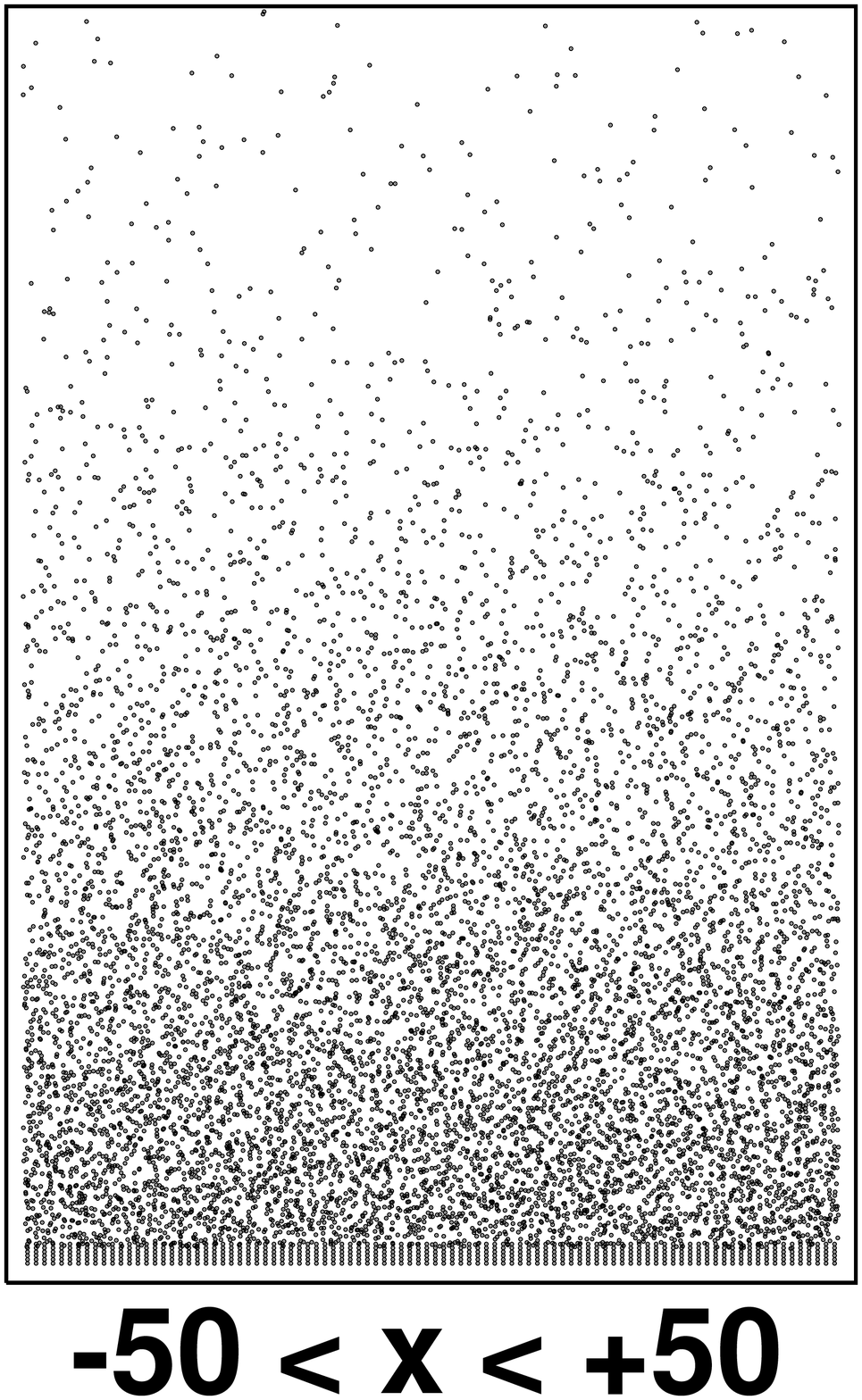}
\caption{
Gravitational isothermal equilibrium at unit temperature for
$n_xn_y = 96 \times 96 = 9216$ moving particles above $6 \times 96 = 576$
boundary particles fixed at the bottom of the system.  The width of the
system is $n_x = 96$. The height is
unbounded. The field strength $g = 4/n_y$ is chosen so that the maximum
density matches that of the fixed particles at the bottom:
$\rho = N/V = 2$ at $y=0$.  This snapshot is typical of a long simulation
used to calculate the smooth-particle pressure profiles shown in
{\bf Figure 2}.  In all of the figures dimensionless (or ``reduced'')
units are used.  These follow from the definitions of unity for the
particle mass, Boltzmann's constant, and the length and energy scales
in the interparticle forces derived from the cusp potential of Sec. II
and the Hooke's-law potential of Sec. IV.
}
\end{figure}

For convenience we have chosen a situation in which the potential and kinetic
parts of the pressure are equally important.  At unit temperature ($kT = 1$)
and a density of 2 ($\rho = Nm/V = N/V = 2$), we have
$$
P^\Phi \simeq (N^2/2V) = \rho ^2/2 = 2 \ ; \ P^K = \rho kT = 2 \ .
$$
We choose the gravitational acceleration $g$ so that the ``weight'' of a column
of unit width and containing $n_y$ particles is equal to the maximum pressure,
$4$, at the maximum density, $\rho (y=0) = 2$.  In this case the mechanical
equilibrium force-balance density and pressure profiles are:
$$
(\rho + 1)(d\rho /dy) = -\rho g \longrightarrow \rho - 2 + \ln (\rho/2) = -gy \ ;
$$
$$
P(y) =  P^\Phi(y) + P^K(y) = g\int_y^\infty \rho(y)dy \ .
$$

We test these analytic results against a molecular dynamics
simulation carried out {\em isothermally}\cite{b4,b5,b6} at a constant
temperature of unity.  At and below the bottom  $y=0$ of the column
we place $6n_x$ boundary particles in an area of $3n_x$ (corresponding to
the maximum density, 2).  See Figure 1.  We also include a short-ranged
repulsive force,
$$
F^{\rm rep}(y<0) \equiv -100y^3 \ ,
$$
which is applied to those few moving particles which occasionally penetrate
the boundary at $y=0$.

With periodic boundaries in $x$ and a repulsive boundary at $y=0$
a 9216-particle simulation gives the typical configuration we showed in
{\bf Figure 1}.  The corresponding kinetic and potential pressure profiles,
averaged vertically with Lucy's one-dimensional weight function,
are compared to the analytic force-balance profile in {\bf Figure 2}.
Evidently the agreement is quite good, and would be qualitatively in
error were the kinetic contribution to the pressure tensor omitted.

\begin{figure}
\includegraphics[height=12cm,width=10cm,angle=-90]{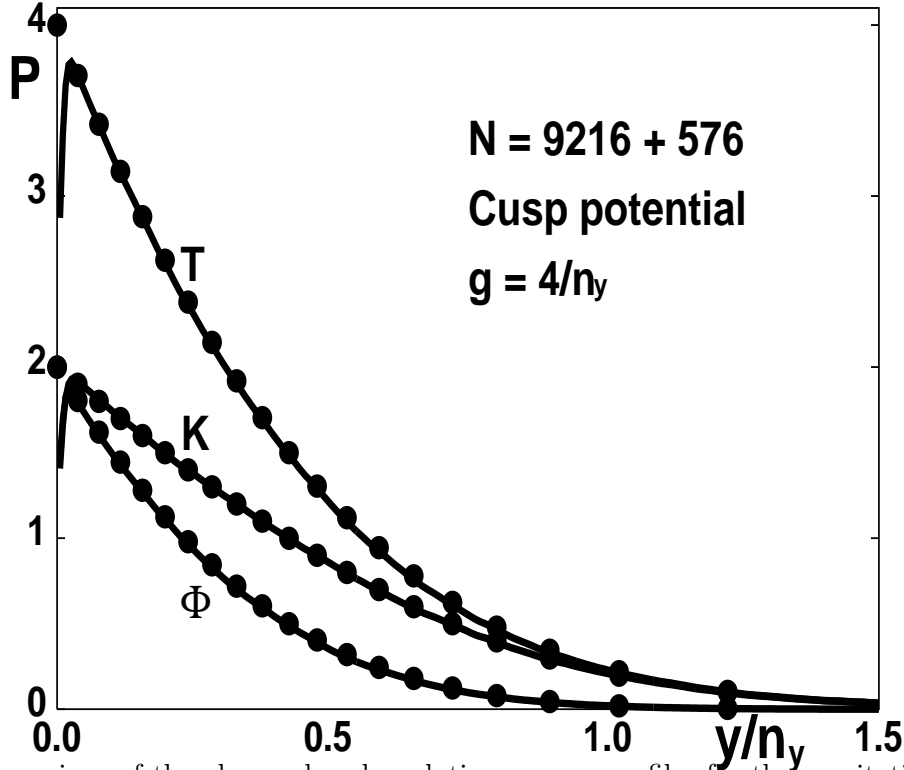}
\caption{
Comparison of the observed and analytic pressure profiles for the
gravitational problem shown in {\bf Figure 1}.  From top to bottom
the three curves are the total ($T$ or Irving-Kirkwood), kinetic
($K$), and potential ($\Phi$ or Zhou) contributions to the pressure
profile.  These observed pressure contributions are calculated as
smooth-particle averages.  The points correspond to the analytic
expressions from the isothermal equation of state
$P_T = P_\Phi + P_K = (\rho ^2/2) + \rho$.
}
\end{figure}

\section{Rotational Equilibration}

Next we consider the influence of the kinetic pressure on the mechanical
equilibrium of a rotating {\em solid}.  We can use molecular dynamics
to determine the thermal (velocity-dependent) properties of an isolated
{\em rotationless} crystal. For this study we have chosen a nearest-neighbor
Hooke's-Law interaction,
$$
\phi_{\rm Hooke} = \frac{\kappa }{2}(|r| - d)^2 \ .
$$
with the force constant $\kappa $, characteristic length $d$, and particle
mass $m$ all set equal to unity.  To make contact with continuum mechanics
we write the stress tensor in terms of the displacement vector $u$ and
elastic constants $\lambda $ and $\eta $:
$$
\sigma = \lambda \nabla \cdot u I + \eta[(\nabla u) + (\nabla u)^t] \ , 
$$
where $I$ is the unit tensor, with $I_{xx} = I_{yy} \equiv 1$ and
$I_{xy} = I_{yx} = 0$.  For
the nearest-neighbor Hooke's-law crystal the Lam\'e constants are known,
$$
\lambda = \eta = \sqrt{3/16}\kappa \ .
$$
as is also the complete vibrational frequency distribution along with
the bulk and surface entropies.  See Chapter 4 of Reference 5 for details.

\begin{figure}
\includegraphics[height=16cm,width=10cm,angle=-90]{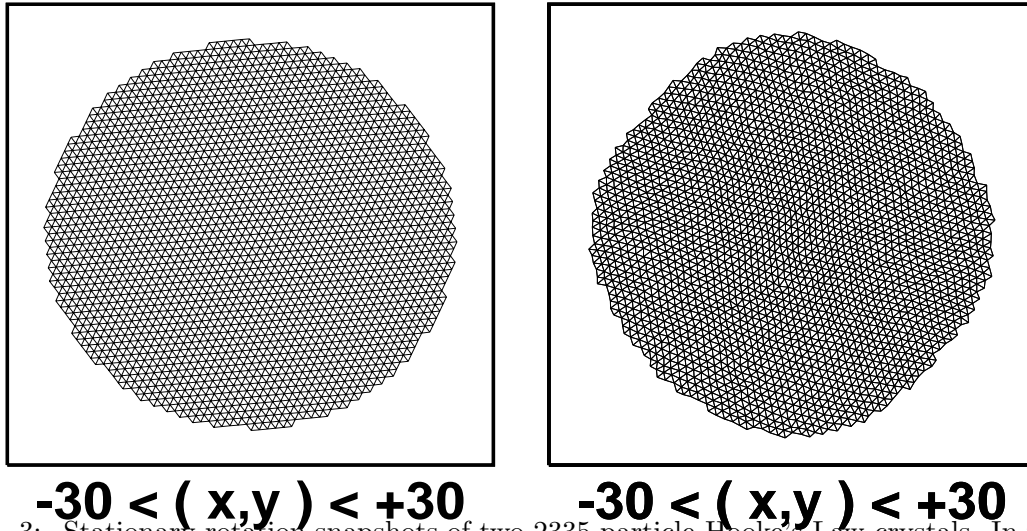}
\caption{
Stationary rotation snapshots of two 2335-particle Hooke's-Law crystals.
In the rotationless stress-free case all 6828 nearest-neighbor
distances are unity.  In the steady rotational situations shown
here, both with an angular frequency $\omega = 0.01$, the tensile strain
offsetting the centrifugal forces is maximized at the center of the
rotating solid.  The left view is a cold solid.  The right view has
a temperature $kT = 0.01$.
}
\end{figure}

The radial displacement in a rotating disk of radius $R$, $u(r)$, as
well as the corresponding stress tensor $\sigma $ are well-known
results of linear elastic theory\cite{b11}.  A derivation for our
two-dimensional situation is sketched in the Appendix.  The results are:
$$
u(r) =  (\omega ^2r/18)[5R^2 - 2r^2] \ ;
$$
$$
\sigma _{rr} = (\rho \omega ^2/12)[5R^2 - 5r^2] \ ; \
\sigma _{\theta \theta} = (\rho \omega ^2/12)[5R^2 - 3r^2] \ .
$$
The stress components satisfy the radial force-balance equation for
a plane-polar-coordinate volume element $rdrd\theta $ rotating at the
angular frequency $\omega $:
$$
+\rho \ddot r = -\rho r\omega ^2 =
(d\sigma _{rr}/dr) + (\sigma _{rr} - \sigma_{\theta \theta})/r  \ .
$$

In the comoving and corotating frame, where stress is the negative of
the momentum flux, rotation provides a centrifugal force per unit mass
varying as $\omega ^2$.

To compare these results from linear elasticity to molecular dynamics
simulations, consider the stationary rotation of a Hooke's-Law lattice.
{\bf Figure 3} shows two nominally stationary states of a 2335-particle
solid with an angular velocity of $\omega = 0.01$. The cold crystal is
shown at the left.  The kinetic temperature of the warm crystal shown
on the right is $kT$ = 0.01.  The 2335-particle crystal is nearly
circular.  It is the smallest with 36 particles equidistant from the
origin (at $\sqrt{637} \simeq 25.239$).  Both these rotational problems
were initialized by thermostating the radial momenta\cite{b4,b5} while
rescaling the angular momenta to generate  thermally-equilibrated
steadily-rotating solid disks.  During the first half of each run two
separate rescaling, or ``Gaussian'', thermostats were applied, so as
to keep the radial temperature and the angular velocity constant.

{\bf Figure 4} illustrates the approximately-quadratic dependence of
the maximum tensile stress on the rate of rotation for small angular
velocities.  For comparison with the simulation results
the linear-elastic stress at the center of a disk with the same mass,
$Nm = 2335$, and a series of rotation rates $\omega $ is also shown.
The agreement is correct to four figures as $\omega \rightarrow 0$.

\begin{figure}
\includegraphics[height=12cm,width=10cm,angle=-90]{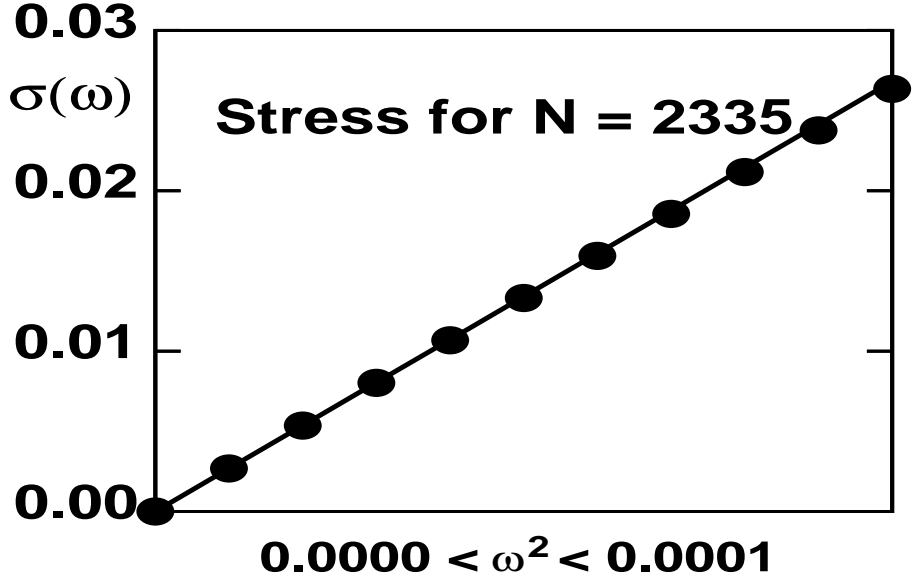}
\caption{
Angular velocity dependence of the cold-crystal maximum tensile stress
on rotation rate.  The molecular dynamics data, shown here as points,
for nearly circular solids of the type shown in Figure 3, agree with the
linear elastic result (shown as a straight line in the figure) for disks
to four figures as the rotation rate goes to zero.  The linear-elastic
result is $\sigma _{\rm max}V/N = (5N\sqrt{3/4}/12\pi)\omega ^2$.
}
\end{figure}

Let us next consider the stresses in a thermally excited rotating
crystal, computed  according to the virial theorem using Irving and
Kirkwood's formulation of the atomistic stresses.  The Hooke's-law
nature of the particle interactions guarantees that our model crystals
will not melt.  But as temperature rises the deformation can become
quite large, so that linear elastic theory no longer applies. Figure 5
is a typical view of a rotating specimen at a rotation rate of
$\omega = 0.01$ and a kinetic temperature (relative to rigid-body
rotation) $kT = 0.02$.

The simplest route to the polar-coordinate stress tensor is, first, to
calculate the kinetic and potential parts of each particle's pressure
tensor in Cartesian coordinates:
$$
(P_{xx}^KV)_i = (p_x^2/m)_i \ ; \
(P_{xy}^KV)_i = (p_xp_y/m)_i \ ; \
(P_{yy}^KV)_i = (p_y^2/m)_i \ ; \
$$
$$
(P_{xx}^\Phi V)_i = \frac{1}{2}\sum_j(xxF/r)_{ij} \ ; \
(P_{xy}^\Phi V)_i = \frac{1}{2}\sum_j(xyF/r)_{ij} \ ; \
(P_{yy}^\Phi V)_i = \frac{1}{2}\sum_j(yyF/r)_{ij} \ .
$$
In keeping with the Irving-Kirkwood picture, the potential contributions
to the pressure tensor are divided evenly between pairs $\{i,j\}$ of
interacting particles.  The polar-coordinate representation for each
particle's pressure tensor follows from the Cartesian representation
by a simple rotation, which can be written as a pair of matrix
multiplications:
$$
(PV)_{\rm polar} = R \cdot (PV)_{\rm Cartesian} \cdot R^t \ ; \
$$
$$
R_i =
\left[
\begin{array}{cc}
+\cos(\theta_i)  &  \ +\sin(\theta_i)  \\
-\sin(\theta_i)  &  \ +\cos(\theta_i)     
\end{array}
\right]
\ ; \ \theta _i = \arctan (y/x)_i \ .
$$

{\bf Figure 5} illustrates a thermally-excited rotating Hooke's-law crystal.
For the figure we have chosen the temperature so that the thermal stresses
make a significant contribution to the pressure tensor.  The radial stress
vanishes at the disk boundary, while the circumferential ``hoop'' stress
remains tensile there in conformity to the predictions of linear elasticity.

\begin{figure}
\includegraphics[height=6cm,width=8cm,angle=-90]{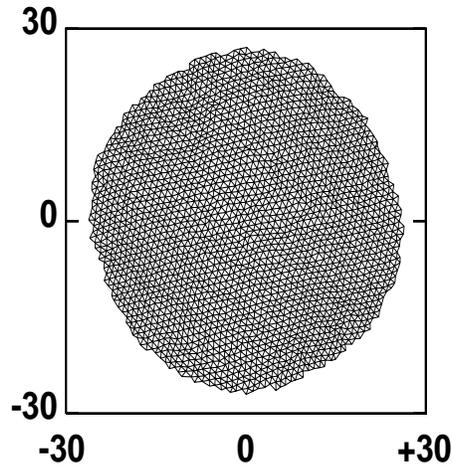}
\caption{
View of a rotating 2335-particle Hooke's-law crystal at an angular
velocity of 0.01 and a temperature $kT = 0.02$.
}
\end{figure}

\begin{figure}
\includegraphics[height=12cm,width=10cm,angle=-90]{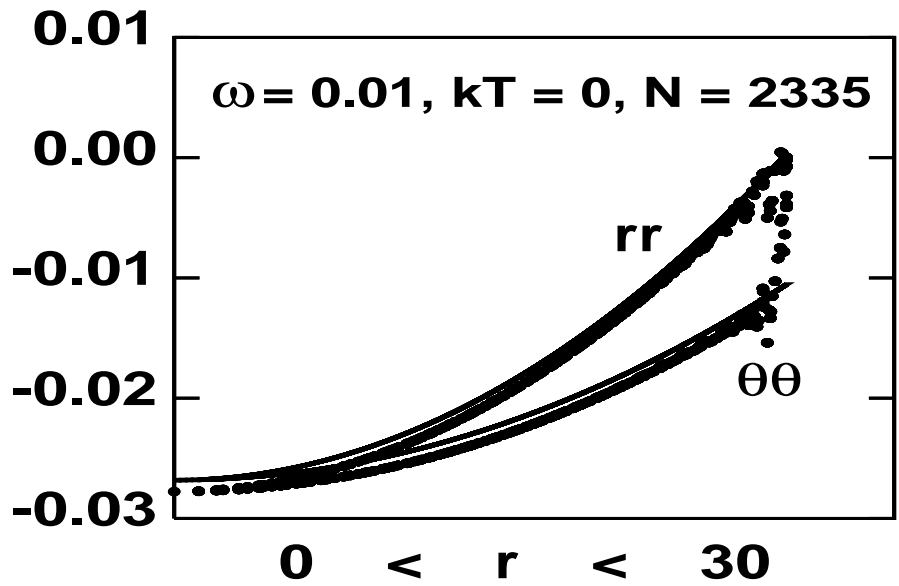}
\caption{
$PV$ in the rotating cold crystal of Figure 3 with $\omega = 0.01$.
The theoretical radial and circumferential components are shown as
lines based on the expressions derived in the Appendix.
}
\end{figure}

The stresses in two rotating crystals, one cold and one hot, are
compared with the theoretical results from elastic theory in {\bf
Figures 6 and 7}.  The agreeement is nearly perfect, and would be
spoiled if the kinetic contributions were not included.  In particular,
omitting the kinetic contribution to the radial stress would be quite
inconsistent with the vanishing of that stress component at the
boundary of the disk.

\begin{figure}
\includegraphics[height=12cm,width=10cm,angle=-90]{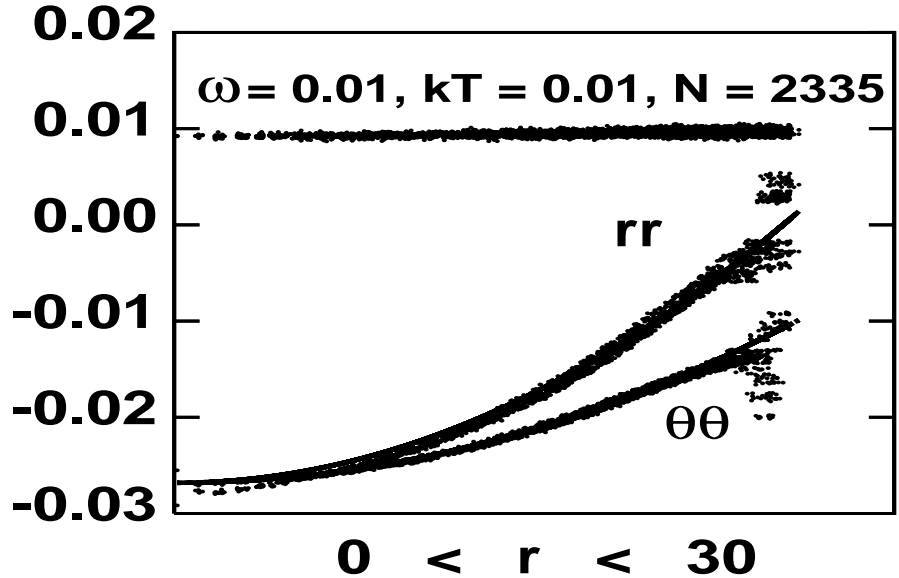}
\caption{
Time-averaged stresses in the warm rotating thermally-excited crystal of
Figure 3 with $\omega = 0.01$ and $kT = 0.01$.  The thermal contributions
to $(PV)_{rr}$ and $(PV)_{\theta \theta}$ are the points at the top.  The
theoretical expressions for the stress (based on the cold-crystal elastic
constant) shown as lines in the Figure agree well with the points
representing results from molecular dynamics.  The molecular dynamics
results include both the potential and kinetic contributions to the
comoving corotating stresses.
}
\end{figure}

\section{Conclusion}

Both the gravitational and the rotational problems show excellent
correspondence between conventional continuum mechanics and atomistic
mechanics {\em provided that both the kinetic and potential parts of the
pressure tensor are included} in the analysis.  Although for stationary
solids the solely potential form for the virial theorem is correct, the
number and type of problems which can be studied numerically is greatly
enhanced by including Irving and Kirkwood's ideas coupled with the
smooth-particle averaging introduced by Lucy and Monaghan in 1977. For
well-defined local properties, both at, and especially away from
equilibrium, it is essential that these properties be measured in a
coordinate frame that moves with the material.  It is no accident that
the fundamental equations of continuum mechanics take their simplest
form in the comoving frame.  In particular, the pressure (or stress)
and temperature tensors, as well as the heat flux, only make sense in
this frame.  Stress and pressure cannot depend upon the chosen coordinate
system.  Hence we must choose the ``comoving'' ``corotating''
``Lagrangian'' frame.  In that frame the pressure tensor is simply the
momentum flux, and has both potential and kinetic contributions, as shown
clearly in the two problems solved here.

\section{Appendix}

The stationary rotation, at angular velocity $\omega $, of an elastic disk
of radius $R$ with equal Lam\'e constants $\lambda = \eta = \sqrt{3/16}$
obeys the force-balance equation in the comoving frame,
$$
0 = +\rho \omega ^2r + (\partial \sigma _{rr}/\partial r) +
(\sigma_{rr} - \sigma _{\theta \theta})/r \ .
$$
This macroscopic problem corresponds to a microscopic model composed of
unit-mass particles linked by nearest-neighbor Hooke's-law springs.  Both
the spring constant and the rest length of the springs are taken equal
to unity.
The displacement responsible for the radial strain
$\epsilon _{rr} = (du_{rr}/dr)$ causes a corresponding strain in the
circumferential direction, $\epsilon _{\theta \theta} = (u/r)$.  The
stresses,
$$
\sigma _{rr} = \eta [3(du/dr) + (u/r)] \ ; \
\sigma _{\theta \theta} = \eta [(du/dr) + 3(u/r)] \ ,
$$
convert the force-balance to an ordinary differential equation:
$$
r^2(d^2u/dr^2) + r(du/dr) -u = -\omega ^2r^3/3 \ ,
$$
with a unique solution such that the radial stress vanishes at $R$:
$$
u(r) = (\rho \omega ^2r/48\eta)[5R^2 - 2r^2] =
                (\omega ^2r/18)[5R^2 - 2r^2]
\ .
$$
This solution can be used to generate the maximum tensile stress in the disk
as well as the stress and strain profiles.
$$
\sigma _{rr} =           (\rho \omega ^2/12)[5R^2 - 5r^2] \ ; \
\sigma _{\theta\theta} = (\rho \omega ^2/12)[5R^2 - 3r^2] \ . \
$$

\end{document}